\documentclass[aps,twocolumn,a4paper,showpacs, superscriptaddress]{revtex4}
\usepackage{graphicx}
\usepackage[all]{xy}
\usepackage{amsmath}
\usepackage{amssymb}
\usepackage{color}
\usepackage[brazil,brazilian]{babel}

\newcommand{\bb}{\bibitem}
\newcommand{\bes}{\begin{subequations}}
\newcommand{\ees}{\end{subequations}}
\def\ben{\begin{eqnarray}}
\def\een{\end{eqnarray}}
\def\be{\begin{equation}}
\def\ee{\end{equation}}
\def\sech{\text{sech}}
\def\sech{\textrm{sech}}

\begin{document}

\title{Scalar fields and defect structures: perturbative procedure for generalized models}
\author{C.A.G. Almeida}
\affiliation{Departamento de Ci\^encias Exatas, Universidade Federal
da Para\'{\i}ba, 58297-000 Rio Tinto, PB, Brazil}
\author{D. Bazeia}
\affiliation{Instituto de F\'\i sica, Universidade de S\~ao Paulo, 05314-970 S\~ao Paulo, SP, Brazil}
\affiliation{Departamento de F\'{\i}sica, Universidade Federal
da Para\'{\i}ba, 58051-970 Jo\~ao Pessoa, PB, Brazil}
\affiliation{Departamento de F\'{\i}sica, Universidade Federal de Campina Grande, 58109-970, Campina Grande, Para\'\i ba, Brazil}\author{L. Losano}
\affiliation{Departamento de F\'{\i}sica, Universidade Federal
da Para\'{\i}ba, 58051-970 Jo\~ao Pessoa, PB, Brazil}
\affiliation{Departamento de F\'{\i}sica, Universidade Federal de Campina Grande, 58109-970, Campina Grande, Para\'\i ba, Brazil}
\author{R. Menezes}
\affiliation{Departamento de Ci\^encias Exatas, Universidade Federal
da Para\'{\i}ba, 58297-000 Rio Tinto, PB, Brazil}
\affiliation{Departamento de F\'{\i}sica, Universidade Federal de Campina Grande, 58109-970, Campina Grande, Para\'\i ba, Brazil}

\date{\today}

\begin{abstract}
We develop a general procedure to deal with defect structures in generalized models, described by a single real scalar field, in $(1,1)$ spacetime dimensions. The models that we consider have the standard kinetic and potential contributions modified to include corrections that depend on a single small parameters, used to control modification on the kinematics and the potential. We start with standard model that engenders stable defect structures, and we show how to obtain new structures for the generalized models. We examine distinct aspects of the new deformed solutions, including linear stability. We work with several distinct modifications, and we show how to make the new defect structures stable, controlled by the parameter that modify the standard theory. We illustrate the procedure with examples of current interest to high energy physics. 

\end{abstract}
\maketitle
\section{Introduction}

Defect structures are nonperturbative finite energy static solutions of classical field theories. They are of current interest to several distinct areas of nonlinear science; see, e.g.,\cite{R,V,K,W,M,Wa,E}. In high energy physics, defect structures appear very naturally through phase transitions in the early universe \cite{V,K}, and can also be used to generate braneworld scenarios \cite{brane,W}. In condensed matter they may, for instance, describe pattern formation \cite{Wa} and properties of magnetic systems \cite{E}. The type of defect as well as its properties depend on the specific details of the symmetry breaking one wants to investigate.

In the simplest case, we consider models describing spontaneous breaking of discrete symmetry. Here the presence of defect structures in models of scalar fields with standard dynamics, are static solutions of the equations of motion. The equations of motion may sometimes be reduced to first-order differential equations \cite{ref9}, with the model being the bosonic portion of more general supersymmetric field theories \cite{S}. One usually refers to this possibility as the first-order framework \cite{blmo07}, where the defect structures that appear as solutions of first-order equations are named BPS states \cite{bps}. Usually, the first-order equations lead to analytic solutions, and an interesting route to this can be implemented through the deformation procedure, as described in Ref.~\cite{deformation}. However, one also finds problems that require numerical calculations; see, e.g.,
Ref.~ \cite{ref12}.

Another interesting route is motivated by applications to cosmology \cite{ref14}, and several distinct modifications of the standard field theory have been investigated. Usually, the generalized models are named k-field models, since they modify the kinematics of the standard field theory, aimed to contribute to understand the accelerated expansion of the Universe. In particular, recent works dealing with specific issues concerning defect structures in generalized k-field models can be found in \cite{ref13}; also, in \cite{ref15} one finds investigations engendering applications to the braneworld scenario with a single extra dimension of infinite extent.

We understand that the search for defect structures in generalized models is much more involved then in the standard situation, because the modification in the kinematics introduces new kinds of nonlinearities, whose study is neither easy nor direct. For this reason, in the present paper we aim to introduce a general formalism to obtain defect structures for generalized field theory, composed by the standard model, enlarged through the addition of a function $F(\phi,X)$, where $X=\frac12 \partial^\mu\phi \partial_\mu \phi$, controlled by a small parameter, $\alpha$, to be used to control the $O(\alpha)$ corrections to the standard theory. Here we consider three distinct possibilities, with the additional contributions modifying the standard kinematics, or the potential, or both the kinematics and the potential.

In the present study, we start reviewing some basic facts about one-field models in Sec.~II. Then, in Sec.~III we introduce the formalism to deal with defect structures in the generalized models, described by standard model, enlarged with new nonlinearities, controlled by the function $F(\phi,X)$. The methodology is inspired in \cite{Almeida:2001pt}, but here we enlarge the scope of the method with the presence of nonlinearities depending on the derivative of the scalar field, an important addition not considered before. We then illustrate the procedure with several distinct examples in Sec.~IV,
and we then finish the work in Sec.~V, where we include some comments and conclusions.
As we show below, up to $O(\alpha)$ one can write interesting general results,
which we believe are of direct interest to high energy physics.
 
\section{Generalities}

Let us start with the standard action, described by one real
scalar field in $(1,1)$ spacetime dimensions. We use natural units, and we shift field and the space and time coordinates to make everything dimensionless. The action can be written in the form
\bes\ben\label{act1}
{S_0}=\int d^2 x \,{\cal L}_0\,,
\\
{\cal L}_0=X - V(\phi)\,,
\\
X=\frac12 \partial_\mu \phi\, \partial^\mu \phi,
\een\ees 
and $V(\phi)$ is the potential, which specifies the model.
The energy-momentum tensor is given by
\be
\label{Tmunu} T^0_{\mu\nu}=\partial_\mu \phi
\partial_\nu \phi - g_{\mu\nu}X + g_{\mu\nu} V(\phi)\,. 
\ee
Since in the search for defect structures we are interested in static configuration, [$\phi=\phi(x)$], we can write the energy $\rho(x)=T_{00}$ and stress $p(x)=T_{11}$ densities as follows
\bes\label{prho} 
\ben
\rho_0(\phi,\phi^\prime)=\frac 12 \phi^{\prime 2} + V(\phi)\,, \\
p_0(\phi,\phi^\prime)=\frac 12 \phi^{\prime 2} - V(\phi)\,, 
\een 
\ees
where the prime stands for derivative with respect to the spatial
coordinate.
The equation of motion for the scalar field is
\be\label{EqofMotion}
\ddot\phi-\phi^{\prime\prime}+\dfrac{dV}{d\phi}=0\,,
\ee
where dot stands for time derivative. For static solutions this equation reduces to
\be\label{eqmo1} 
\phi^{\prime\prime} = V_\phi\,,
\ee
where $V_\phi=dV /d\phi$.
The simplest solution for this equation is when the field is
homogeneous, which are the critical points of the potential, so that
$V_\phi=0$.
Now, if we write the potential in the following form 
\be\label{potential}
V(\phi)=\frac 12 W_\phi^2\,,
\ee 
the minima of the potential are the critical points of the superpotential, $W=W(\phi)$.
The stress density can be written as 
\be
p_0(\phi,\phi^\prime)=(1/2)(\phi^\prime - W_\phi)(\phi^\prime + W_\phi)\,,
\ee
We make the stress density vanish, leading to the first order equations
\be\label{first-order} 
\phi^\prime =\pm W_\phi,
\ee 
which solve the equation of motion (\ref{eqmo1}). Solutions of the first-order equations are named BPS states \cite{bps}.

In order to circumvent problems with unstable solutions, we assume
that $W=W(\phi)$ is a smooth function. For convenience, in the
first order equations, we include $\pm$ into the definition
of $W$, since the sign of $W$ is not seen by the potential; thus,
we write $\phi^\prime=W_\phi$. Once again, we note that the global minima of the
potential ($v_{i}, i=1,2,\ldots$) are the critical points of the
superpotential, and this implies that the concavity of the potential at each
minimum is always positive.
We can integrate the energy density $\rho(x)$ to get the total 
energy of the solution in the form
$E_0=|W(\phi(\infty))-W(\phi(-\infty))|$.
Let us consider  topological solution whose asymptotic behavior
is constant and reproduces a minimum of the potential, such that
\be
\frac{d\phi}{dx} \to 0 \,,\,\,\,\,\, {\rm for} \,\,\,\,x\to \pm
\infty 
\ee
The topological features of the solutions can be seem from the topological current $j^\mu=\epsilon^{\mu\nu} \partial_\nu \phi$ \cite{R}.
For static solutions the topological charge is 
\be\label{tcharge}
Q=\int^\infty_{-\infty} dx j^0 = \phi(\infty) -\phi(-\infty)\,.
\ee
It only depends on the asymptotic values of the solution.

In order to examine linear stability of the static solutions $\phi=\phi(x)$, we take
\be\label{eta123}
\phi(x,t)=\phi(x)+\eta(x,t)\,,
\ee
where $\eta(x,t)$ is small perturbation around the static solution.
The first-order contribution coming from Eq.~\eqref{EqofMotion} is $\partial^\mu\partial_\mu \eta +
V_{\phi\phi}\eta =0$. For static solutions, we can choose
$\eta(x,t)=\eta(x)\cos(\omega\, t)$, leading to 
\be
-\eta^{\prime\prime} + U(x) \eta = \omega^2 \eta 
\ee 
where $U(x)=V_{\phi\phi}$ is the quantum-mechanical like
potential. This equation can be factorized to 
\be \left(\frac{d}{dx}
+ W_\phi\right)\left(-\frac{d}{dx} + W_\phi\right)\eta = \omega^2\eta\,, 
\ee 
whose zero mode is $\eta_0 \propto \phi^\prime$, showing that the static solutions are stable.

An important example is the $\phi^4$ model, engendering spontaneous symmetry breaking. It is defined by the potential
\be\label{phi4} 
V(\phi)=\frac12 (1-\phi^2)^2\,,
\ee
generated by the superpotential
$
W(\phi)=\phi-\frac13 \phi^3\,.
$
The BPS solutions connecting the two minima at $\phi=\pm1$ are 
$
\phi_0(x)=\pm \tanh(x)\,,
$
which have energy density 
$
\rho_0(x)=\sech^4(x)\,
$
and energy $E_0=4/3$. In this case, the quantum-mechanical potential is given by
\be\label{uphi4}
U(x)=4-6\,\sech^2(x)\,,
\ee
and the (normalized) zero mode is
$
\eta_0(x)=\sqrt{{3}/{4}}\;\;\sech^2(x)\,.
$
This potential has the zero mode and another bound state.

Another relevant example is the sine-Gordon model. Here, the
potential is given by 
\be\label{vsg} 
V(\phi)=\frac12 \cos^2(\phi)\,.
\ee
The superpotential is given by $W(\phi)=\sin(\phi)$.
It has equidistant minima at $\phi_{min}=(n+1/2)\pi$ and maxima at $\phi_{max}=n\pi$, with$n=0,\pm1,\pm2,...$ The BPS solutions are 
\be\label{solsg}
\phi^k_0(x)=\arcsin(\tanh(x))+k\pi,\;\;\;\;k=0\pm1,\pm2,...,
\ee 
with the same energy density
$
\rho_0(x)=\sech^2(x)\,,
$
and total energy $E_0=2$.
The quantum-mechanical potential has the form
\be\label{usg}
U(x)=1-2\,\sech^2(x)\,,
\ee
and the zero mode is $\eta_0=\sqrt{1/2}\,\,\sech(x)$. Here the zero mode is the only bound state.

\section{NEW MODELS}

Let us now consider new models, introduced by the addition of a new term to the standard action presented in the previous Sec.~II. The new action has the form
\be\label{action1222} 
S=S_0 + \alpha \int d^2x \,F (\phi, X)\,, 
\ee 
where $S_0$ is the unperturbed action given by Eq.~\eqref{act1}, $F(\phi,X)$ is in principle an arbitrary function of $\phi$ and $X$, and $\alpha$ is a very small parameter,
used to control the power of the perturbative expansion.
This procedure was already considered in the work \cite{Almeida:2001pt}, 
but here $F$ includes dependence on $X$, meaning that the extra term may contain nonlinearities depending on the derivative of the scalar field, a possibility that has remained unexplored until now.
 
The energy-momentum tensor for the new model can be written as 
\be
{T}_{\mu\nu}=T^0_{\mu\nu} + \alpha \left(F_X
\partial_\mu \phi \partial_\nu \phi - g_{\mu\nu} F (\phi, X)\right)\,,
\ee 
with $T^0_{\mu\nu}$ given by Eq.~\eqref{Tmunu}. 
For static solutions, the energy and stress densities can be written as 
\bes\label{energyd2}
\ben
{\rho}(\phi,\phi^\prime)&=&\rho_0(\phi,\phi^\prime) -\alpha\, \rho_{\alpha}
\\
{p}(\phi,\phi^\prime)&=&p_0(\phi,\phi^\prime) + \alpha\,p_{\alpha}
\\
\rho_{\alpha}&=&F(\phi,X)
\\
p_{\alpha}&=&F-2F_{X}X\een 
\ees 
where $\rho_0$ and $p_0$ are given by Eq.~\eqref{prho}. The equation of motion for the new model \eqref{action1222} has the form 
\be
\partial^\mu \partial_\mu \phi + W_\phi W_{\phi\phi} = \alpha \left(F_\phi-\partial_\mu (F_X \partial^\mu \phi)\right)\,.
\ee
In the case of static solutions, we can write
\be\label{qe}
-\phi^{\prime\prime}+ W_\phi W_{\phi\phi} = \alpha (F_\phi+(F_X \phi^\prime)^\prime)\,.
\ee
Here, we consider the solution in the form 
\be\label{pphi}
\phi(x)=\phi_0(x) + \alpha \phi_\alpha(x) \,,
\ee
where $\phi_0(x)$ is the static solution when $\alpha$ vanishes. Expanding the Eq.~\eqref{qe} in terms of $\alpha$, for $\alpha$ very small, up to first-order in $\alpha$
we obtain
\be 
-\phi_\alpha^{\prime\prime} + (W_\phi
W_{\phi\phi\phi}+W_{\phi\phi}^2) \phi_\alpha=  F_\phi+(F_X
\phi^\prime)^\prime\,,
\ee
where $\phi=\phi_0$, is the homogeneous solution when $\alpha$ vanishes. In the generalized model, the homogeneous solution is given by, up to first-order in $\alpha$
\be\label{Criticalpoints} 
\phi=\phi_0 + \alpha\left(\frac{F_\phi}{W_\phi
W_{\phi\phi\phi}+W_{\phi\phi}^2}\right)_{\phi=\phi_0}.
\ee
This result shows that the asymptotic limit of the topological
solution may change only when $F$ depends on $\phi$.
The equation of motion \eqref{qe} can be integrated to
give constant stress density, $\tilde{p}=c$. Taking $c=0$, we can write the first-order equation for static field
\be\label{first2}
\phi^{\prime} = \sqrt{W_\phi^2 - 2\alpha p_\alpha}\,.
\ee
If we now use \eqref{pphi} and \eqref{first-order}, we obtain
\be 
\phi_\alpha^{\prime} = W_{\phi\phi}(\phi_0)\; \phi_\alpha -
\frac{p_\alpha(\phi_0,\phi^\prime_0)}{W_\phi(\phi_0)}\,,
\ee
which is solved by
\be\label{alpha}
\phi_\alpha(x)=- \phi_0^{\prime}(x) \int^{\phi_0}_0
\frac{p_\alpha(\phi)}{W_\phi^3} \,\,d\phi\,,
\ee 
after discarding an integration constant, which only induces a
translation of the unperturbed solution.

Let us calculate the contribution to the energy of the perturbed static solution \eqref{pphi}, which arises from the first-order contribution to the energy density, Eq.~\eqref{alpha}. By means of an integration by parts and using \eqref{eqmo1},  we can write (\ref{energyd2}a) as 
\be\label{energydensity2}
{\rho}=\phi^{\prime\;2}_0-\alpha F(\phi_0,\phi^\prime_0)+
\alpha (\phi^\prime_0 \phi_\alpha)^\prime 
\ee 
The expansion of the
energy in power of $\alpha$ can be represented as $E=E_0+\alpha E^{(1)} + \alpha^2
E^{(2)}$+...; thus, the first-order correction has the form 
\be
E^{(1)}=-\int_{-\infty}^{\infty} dx F(\phi_0,\phi^\prime_0)+
(\phi^\prime_0 \phi_\alpha)\Big|_{x\to -\infty}^{x\to \infty} 
\ee 
Particularly, for kinklike solution $\phi_0^\prime \to 0$ when $x\to\pm
\infty$; therefore, the first-order correction to the energy does not depend on the
first-order correction to the solution, resulting in 
\be\label{FirstOrder}
E^{(1)}=-\int_{\phi_0(-\infty)}^{\phi_0(\infty)} d\phi
\,\frac{F(\phi,W_\phi)}{W_\phi}\,.
\ee
This is  a general result, valid for $F=F(\phi,X)$ and for static solution $\phi=\phi(x)$.
We can also show that the $n$th-order correction in $\alpha$ to the energy depends on the
$(n-1)$th-order correction to the field configuration. Thus, the second-order correction to the energy, that depends on $\phi_\alpha$, is given by
\ben\label{SecondOrder}
E^{(2)}=\int_{\phi_0(-\infty)}^{\phi_0(\infty)} \frac{d\phi}{W_{\phi}}\, \Big(\frac{\phi_\alpha^{\prime\,2}}{2}+\frac12U_0(\phi)\phi_\alpha^2&&\nonumber\\
- F_\phi \phi_\alpha + F_X W_\phi \phi_\alpha^\prime \Big)\,,&&
\een
where
\be\label{uz}
U_0(\phi)=W_{\phi\phi}^2+W_\phi W_{\phi\phi\phi}\,.
\ee
The topological charge of the static solution \eqref{pphi} is defined by \eqref{tcharge}, and can be written as
\be Q=Q_0 + \alpha\;\left[\phi_\alpha(\infty) - \phi_\alpha(-\infty)\right]\,,
\ee
where $Q_0=\phi_0(\infty)-\phi_0(-\infty)$ is the topological charge of the unperturbed solution $\phi_0(x)$.

We now focus attention on the linear stability of the defect structures constructed perturbatively, up to the first-order power in $\alpha$. For static solution, using \eqref{eta123} and following the procedure developed in Refs.~\cite{blmo07,Bazeia:2008tj}, we can write
\ben\label{eqof23}
\partial_{\mu}\partial^{\mu}\eta+V_{\phi\phi}\eta=\alpha \Big[-F_X \ddot\eta +((F_{X}+2XF_{XX})\eta^{\prime})^{\prime}&&\nonumber\\ 
+F_{\phi\phi}\eta+(F_{X\phi}\phi^{\prime})^{\prime}\eta\Big].\;\;\;&&\nonumber\\
\;
\een
Now, we consider $\eta(x,t)=\eta(x)\cos(\omega t)$; after substituting this into \eqref{eqof23} we get
\ben
-\eta^{\prime\prime}+V_{\phi\phi} \eta -\omega^2 \eta 
= \alpha \Big[\left((F_X+2F_{XX}X)\eta^\prime\right)^\prime &&\nonumber\\
\!+\left( F_X \omega^2+F_{\phi\phi}+(F_{X\phi}\phi^\prime)^\prime \right)\eta\Big].&&
\een
We can rewrite this equation as a Schr\"odingerlike equation, in the form
\be
\left[-\frac{d^2}{dz^2}+\tilde{U}(z)\right]\tilde\eta=\tilde{\omega}^2 \tilde\eta\,.
\ee
To do this, we follow the procedure described before in Ref.~\cite{blmo07}. We change variables according to
\bes\label{Trans1}
\be
dx =\left(1+\alpha F_{XX}X\right) dz\,,
\ee
and
\be
\eta=\left(1-\frac\alpha2 \left(F_{XX}X+F_X\right)\right) \tilde\eta\,.
\ee
\ees
The above modifications lead us to the quantum-mechanical potential $\tilde{U}(z)$, up to first-order in $\alpha$,
\be\label{uqpert}
\tilde U(z)=U_0(z)+\alpha\; U_\alpha(z)\,,
\ee
where $U_0$ is given by \eqref{uz}, and
\ben\label{ualpha}
U_\alpha(z)&=&\frac 12 (F_X + F_{XX} X)_{zz}  - (F_{X\phi} W_{\phi})_z
- F_X U_0\nonumber\\ 
&&-F_{\phi\phi}+\frac12\frac{dU_0}{dz}\int dz \left(\frac{F}{X}
- 2 F_X+2 F_{XX}X\right),\nonumber\\
&&
\een
with $\phi=\phi_0(z)$. In this case, the ground state corresponding to the potential
\eqref{uqpert} can be written as
\ben
\tilde\eta_0(z)&=&\eta_0(z)+\frac{\alpha}2\,\eta_0(z)\,\left(F_{XX}X+F_X\right)\nonumber\\
&&+\alpha\;\eta_0^{\prime}\int dz\;F_{XX}X\,,
\een
where $\eta_0(z)$ is the zero mode of the standard model, with potential $U_0$.

In the standard theory, the quantum-mechanical potential $U_0$ has the zero-mode $\eta_0 \propto \phi_0^\prime$; this is the ground-state, and ensures stability of the defect structure. Therefore, the correction to the zero-mode energy in the generalized model can be written as
\be\label{deltaw}
\tilde{\omega}_0=\alpha\int^{\infty}_{-\infty} dz\; \eta_0^2(z)\; U_{\alpha}(z)\,,
\ee
Thus, when it is non-negative, the static solutions \eqref{pphi} is stable too.
This can be done by appropriately choosing the sign of $\alpha$.
Alternatively, we can guess stability from the plots of the two potentials,
$U_0$ and $U_{\alpha}$, and this will become clearer below, where
we investigate distinct generalizations, governed by $F(\phi,X)$.

\section{Examples}

Let us now study three distinct generalizations, the first depending only on the scalar  field, with $F(\phi,X)$ changed to $F(\phi)$; the second, depending only on the derivative of the scalar field, with $F(\phi,X)$ changed to $F(X)$; and the third, for $F(\phi,X)$ depending on both the field and its derivative.

\subsection{The case of $F(\phi)$}

Let us investigate the case where $F= F(\phi)$,
with the generalization adding extra terms to the potential. Thus, we call
$F(\phi)=- V_\alpha(\phi)$, which allows writing the potential in the form
\be\label{vpertub}
\tilde V(\phi)=\frac12\left(W_\phi+\alpha
\frac{V_\alpha(\phi)}{W_\phi}\right)^2\,,
\ee
valid up to first-order in $\alpha$. In the range determined by the unperturbed static solution, that is, for $\phi_0(-\infty)\leq\phi\leq\phi_0(\infty)$, from \eqref{Criticalpoints} we see that the minima and maxima are given by
\bes\label{minmax}
\ben
\phi_{min}&=&\phi^0_{min}- \alpha\left(\frac{V_{\alpha,\phi}}
{W_{\phi\phi}^2}\right)_{\phi=\phi^0_{min}}\,,\\
\phi_{max}&=&\phi^0_{max}-\alpha
\left(\frac{V_{\alpha,\phi}}
{W_{\phi}W_{\phi\phi\phi}}\right)_{\phi=\phi^0_{max}}\,,
\een 
\ees
where $\phi^0_{min}$ and $\phi^0_{max}$ are the minima and maxima of the standard potential \eqref{potential}, respectively. The maxima values of the potential are 
\be\label{vmax}
V_{max}=V^0_{max}
\left(1+2\alpha\frac{V_\alpha}{W_\phi^2}\right)_{\phi=\phi^0_{max}}\,,
\ee
where $V^0_{max}$ is a maximum of the standard potential.
Here $p_\alpha=-V_\alpha(\phi)$, and using \eqref{alpha} we have
\be\label{pertphi}
\phi_\alpha(x)=g(\phi_0)\,,
\ee
where
\be
g(\phi)=W_\phi \int_0^{\phi} \frac{V_\alpha(\phi)}{W_\phi^3} \,\,d\phi\,.
\ee
The energy density \eqref{energydensity2} takes the form
\be\label{rhog}
\tilde\rho=\rho_0 (1+ 2\alpha \,g_{\phi}(\phi_0))\,, 
\ee
where $\rho_0=\phi_0^{\prime\;2}$ is the energy density of the standard model, which furnishes the first-order correction to the energy
\be\label{EE1}
E^{(1)}=2\int_{\phi_0(-\infty)}^{\phi_0(\infty)} d\phi \,W_\phi g_\phi\,.
\ee
If we define the superpotential of the generalized model as
\be\label{tw} 
\tilde W(\phi)=W(\phi)+\alpha \int_0^\phi d \phi
\frac{V_{\alpha}(\phi)}{W_{\phi}}\,,
\ee
it is easy to show that 
\be\label{deltaW}
E=E_0+\alpha E^{(1)}=|\tilde W(\phi_0(\infty))-\tilde W(\phi_0(-\infty))|\,.
\ee
In this case, the correction to the quantum-mechanical potential \eqref{ualpha}
can be written in the form
\be\label{uquant2}
U_{\alpha}(x)=3\;W_{\phi}W_{\phi\phi}g_{\phi\phi}+W^2_{\phi}g_{\phi\phi\phi}\,
\ee
where $\phi=\phi_0(x)$. The correction to the zero mode \eqref{deltaw} of the standard model is given by
\be
\tilde\omega_0=\alpha\int^{\phi_0(\infty)}_{\phi_0(-\infty)} d\phi\;W_{\phi} \; U_{\alpha}\,.
\ee
However, we can show that
\ben
\int^{\phi_0(\infty)}_{\phi_0(-\infty)} d\phi\;(W^3_{\phi}g_{\phi\phi\phi}
+3\;W^2_{\phi}W_{\phi\phi}g_{\phi\phi})=0\,,
\een
which leads to $\tilde\omega_0=0$. Therefore, the defect structures of the generalized model are stable, as they are in the standard theory.

Let us now illustrate the procedure with some explicit examples.
Note that, if we choose the perturbation in the form
$V_\alpha=-W_\phi^2/2$, we get the trivial case, leading to
$\label{trivial2} \phi_\alpha=-\frac{1}{2}\,x
\phi_0^{\prime}$.

\begin{figure}[ht]
\includegraphics[scale=0.42]{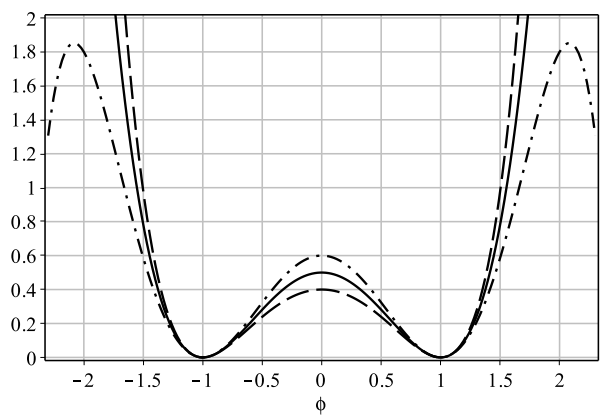} 
\caption{The $\phi^4$ potential \eqref{phi4} (solid line) and the generalized potential \eqref{phi41}  for $\alpha=-0.1$ (dashed line), and for $\alpha=0.1$ (dashed-dotted line).\label{figu1}}
\end{figure}

An interesting choice is given by
\be
\label{Wp3} V_\alpha(\phi)=W_{\phi}^3\,,
\ee
which gives $\phi_{\alpha}=\phi_0 \phi_0^{\prime}$. The mimima and maxima do not change, only the heights of the maxima have a shift given by 
$
\Delta V_{max}=2\alpha V_{max}W_{\phi}(\phi^0_{max})\,,
$
for $\phi_0(-\infty)\leq\phi\leq\phi_0(\infty)$.
From  \eqref{rhog}, the correction to the energy density is
\be
\rho_{\alpha}=2\alpha(\phi_0^{\prime\;3}+\phi_0\phi_0^{\prime\;2}W_{\phi\phi}(\phi_0))\,,
\ee
and from \eqref{EE1} the first-order correction of the energy is
\be
E^{(1)}=\int_{-\infty}^{\infty} dx \phi_0^{\prime\;3}\,.
\ee
For instance, we take the perturbation (\ref{Wp3}) in the $\phi^4$ potential \eqref{phi4}.
In this case, we have the generalized potential
\be\label{phi41}
\tilde V(\phi) = \frac12 \left(1+2\alpha(1-\phi^2)\right)(1-\phi^2)^2\,,
\ee
which was investigated in Refs.~\cite{Christ:1975,Bazeia:2005}. Note that, the maximum at the origin is shifted by $\Delta V_{max}=\alpha$.
For $\phi^2>>1$, the term in $\alpha$ grows, and can not be treated as a small correction. However, we want to find corrections of the defect structures, and they obey the condition $\phi^2<1$, thus legitimating the above procedure. In Fig.~\ref{figu1} we depict some potentials. 

In the current case, the static solution is
\be
\phi(x)=\tanh(x)(1+\alpha\,\sech^2(x))\,,
\ee 
and the correction of the energy is $E^{(1)}={16}/{15}$.
The perturbation of the quantum potential \eqref{uquant2} becomes
\be
U_{\alpha}(x)=36\;\sech^2(x)-42\;\sech^4(x)\,.
\ee

\begin{figure}[ht]
\includegraphics[scale=0.42]{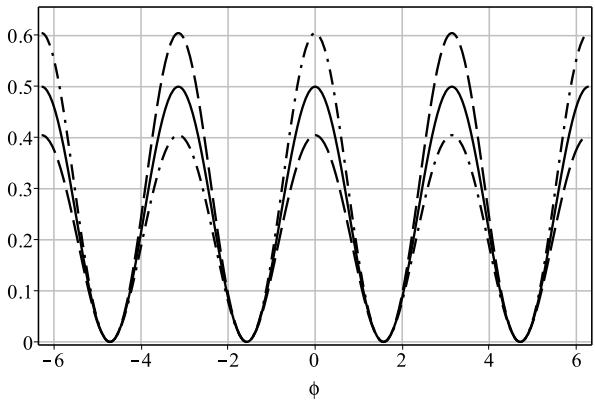} 
\caption{The sine-Gordon potential \eqref{vsg} (solid line) and the generalized potential \eqref{vsg1} for $\alpha=-0.1$ (dashed line), and for $\alpha=0.1$ (dashed-dotted line).\label{figu2}}
\end{figure}

As another example, let us take the perturbation (\ref{Wp3}) in the sine-Gordon potential \eqref{vsg}. It leads to the generalized potential
\be\label{vsg1}
\tilde V=\frac12\left(\cos(\phi)+\alpha\cos^2(\phi)\right)^2\,,
\ee
which is depicted in Fig.~\ref{figu2} for $\alpha=-0.1$. Note that, the maxima at $\phi=n\pi$ are shifted by $\Delta V_{max}=(-1)^n \alpha$. Thus, the generalized potential is a double sine-Gordon potential. 

The static solution is
\be
\phi(x)=\left(1+\alpha\,\sech(x)\right)\phi_0(x)\,,
\ee 
where $\phi_0(x)$ is given by \eqref{solsg}. The corresponding energy is $E=2+\frac{\pi}{2}\alpha$, and the quantum potential \eqref{uquant2} is
\ben
U_{\alpha}(x)&=&6\;\sech(x)-9\;\sech(x)^3\nonumber\\
&+&4\;\tanh(x)\;\sech(x)^2\arcsin(\tanh(x))\,.
\een

\begin{figure}[ht]
\includegraphics[scale=0.42]{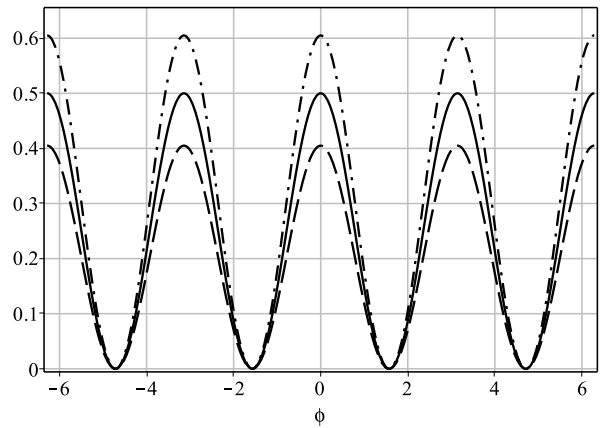}
\caption{The sine-Gordon potential \eqref{vsg} (solid line) and the generalized potential \eqref{vsg2} for $s=1$, with $\alpha=-0.1$ (dashed line) and with $\alpha=0.1$ (dashed-dotted line).\label{figu3}}
\end{figure}

We consider a more general example, applying the perturbation  $V_\alpha=\cos(\phi)\cos(\phi/s)$, 
parametrized by the real parameter $s$, in the sine-Gordon model \eqref{vsg}. Here we get
\be\label{vsg2}
\tilde V=\frac12\left(\cos(\phi)+\alpha\cos\left(\frac{\phi}{s}\right)\right)^2\,.
\ee
The minima and maxima \eqref{minmax} are
\bes\label{minmax2}
\ben\label{min2}
\phi_{min}&=&\frac{(2n+1)\pi}{2}+(-1)^n\cos\left(\frac{(2n+1)\pi}{2s}\right)\\
\phi_{max}&=&n\pi-\frac{(-1)^n}{s}\sin\left(\frac{n\pi}{s}\right)\,,
\een 
\ees
where $n=0,\pm1,\pm2,...$, and from \eqref{vmax} the height of the maxima are
\be
\label{vmax2}
\tilde{V}_{max}=\frac12\left(1+2(-1)^n\alpha\cos\left(\frac{n\pi}{s}\right)\right)\,.
\ee
Then, from \eqref{vmax2} we obtain the multiplicity of the sine-Gordon model, determined by the number of different heights of their maxima.
In general, there is no closed formula for the modification of the static solution \eqref{pertphi}, which  can be calculated for some values of $s$, as we show below. However, the superpotential \eqref{tw} can be written as
\be
\tilde{W}=\sin(\phi)+\alpha\;s\;\sin\left(\frac{\phi}{s} \right)\,,
\ee
and the minima \eqref{min2}, can be used to calculated the energy of the defect structures, straightforwardly.
Note that, the parameter $s$ introduces a new family of sine-Gordon models defined by \eqref{vsg2}, which includes the double and triple sine-Gordon models. For instance, for $s=1$ we have  a sine-Gordon potential with the height of maxima: $\tilde{V}_{max}=(1+2\alpha)/2$, as shown in Fig.~\ref{figu3}.

\begin{figure}[ht]
\includegraphics[scale=0.42]{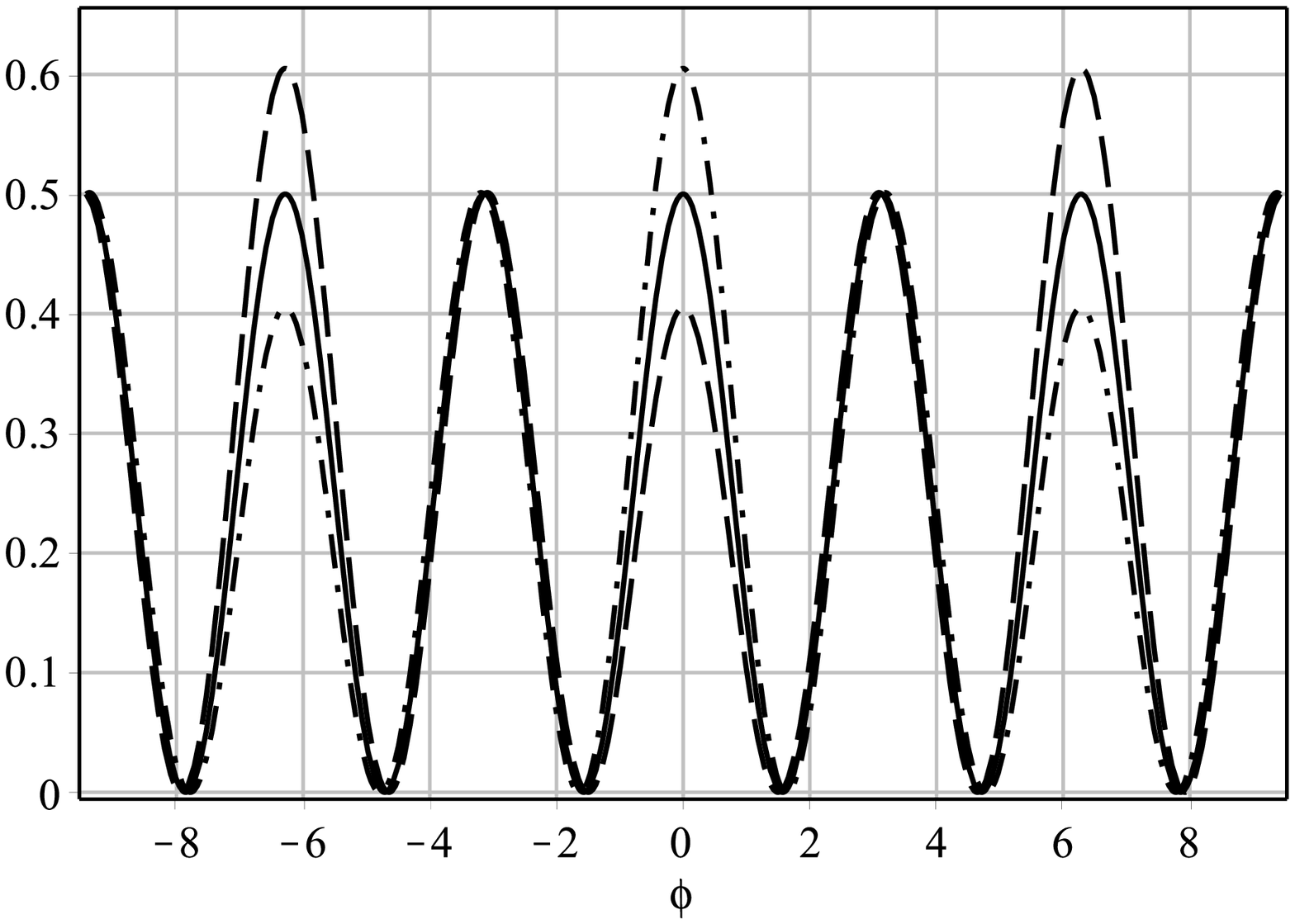} 
\caption{The sine-Gordon potential \eqref{vsg} (solid line) and the generalized potential \eqref{vsg2} for $s=2 $, with $\alpha=-0.1$ (dashed line) and with $\alpha=0.1$ (dashed-dotted line).\label{figu4}}
\end{figure}

In this case, the static solutions are
\be
\phi(x)=\phi_0(x)+\alpha x\;\sech(x)\,,
\ee
where $\phi_0(x)$ is given by \eqref{solsg}, the energy is $E=2+2\alpha$, and the quantum perturbation \eqref{uquant2} is 
\be
U_{\alpha}(x)=2-4\;\sech(x)^2\left(1-x\;\tanh(x)\right)\,.
\ee
For $s=2$, we have a triple sine-Gordon potential with the heights of maxima: $\tilde{V}_{max}=(1-2\alpha)/2$, $1/2$, and $(1+2\alpha)/2$,  as illustrated in Fig.~\ref{figu4}.

The static solutions are 
\be
\phi(x)=\phi_0(x)+\frac{\alpha}{\sqrt{2}}\left(\chi(x)+\sech(x){\rm arctanh}(\chi(x))\right)\,,
\ee
where $\phi_0(x)$ is given by \eqref{solsg} and $\chi(x)=(1-\sech(x))^{1/2}$, the energy is $E=2+2\alpha$, and the perturbation in the quantum potential \eqref{uquant2} is 
\ben
U_{\alpha}(x)&=&\frac{\tanh(x)}{\sqrt{2}\;\chi(x)}\Big(1+\frac74\sech(x)-4\;\sech^2(x)\Big)\nonumber\\
&+&2\sqrt{2}\tanh(x)\;\sech^2(x)\;{\rm arctanh}(\chi(x))\,.
\een

In particular, when $\phi_\alpha=g(\phi)$ is invertible, an alternative way to implement the above generalization is by means of the deformation procedure \cite{deformation}, employing the deformation function $f(\phi)=\phi-\alpha g(\phi)$, as treated in Ref.~\cite{alpha-sg}. For the above model, other values of $s$ are admissible, integer or half-integer,
leading to other sine-Gordon-like models; for instance, for $s=1/2$ we get to the double sine-Gordon model, etc. Detailed investigations can be implemented at will, following the steps given above.

\subsection{The case $F(X)$}

Here we move on to the case where $F=F(X)$, with the generalization depending only on the derivative of the scalar field. We see that, for static solutions,
\begin{equation}
X=\frac12\partial_\mu\phi\partial^\mu\phi=-\frac 12 \phi^{\prime 2},
\end{equation}
and we can use Eq.~\eqref{first-order} to write 
\begin{equation}
X=-\frac12 W_\phi^2\,.
\end{equation}
To deal with a specific example, let us choose the monomial function of $X$ as 
\be\label{kpert1}
F(X)=\frac{2^{n-1}}{n}X|X|^{n-1}\,,
\ee where $n$ is a integer positive
parameter. Using Eq.~\eqref{Criticalpoints}, we see that the homogeneous solutions are the same solutions of the standard model.
From Eq.~(\ref{energyd2}b), the addition contribution for the stress density becomes
\be
p_\alpha(W_\phi)=\frac{2n-1}{2n}W_\phi^{2n}\,,
\ee
whence the perturbation $\phi_\alpha$ given by Eq \eqref{alpha} becomes
\be\label{phia1}
\phi_\alpha=- \frac{2n-1}{2n} \phi_0^{\prime}\;\int^{x}_0 dx\,\phi_0^{\prime\;2n-2}\,,
\ee
which, using \eqref{energydensity2}, gives the energy density
\be\label{energydensity3}
{\rho}=\phi^{\prime\;2}_0+\frac{\alpha}{2n}\phi^{\prime\;2n}_0+
\alpha [\phi^\prime_0 \phi_\alpha]^\prime\,.
\ee 
We use \eqref{FirstOrder} to calculate the first-order correction of the energy by means of
\be\label{E11}
E^{(1)}=\frac{1}{2n}
\int_{-\infty}^{\infty} dx \,\phi^{\prime\;2n}_0\,.
\ee
For instance, for $n=1$, we obtain
\be\label{trivial1}
\phi_\alpha=-\frac{1}{2}\,x\;\phi^\prime_0\,,
\ee
the energy density
\be
\tilde\rho=\phi^{\prime\;2}_0 -\alpha\;x\phi^{\prime}_0\phi^{\prime\prime}_0\,,
\ee
and the energy
\be\label{eny}
E=E_0\left(1+\frac12\alpha\right)\,,
\ee
where $E_0$ is the energy of the defect structure of the standard model.
Also, for $n=2$, Eq. \eqref{phia1} yields
\be\label{expre12} 
\phi_\alpha=- \frac{3}{4} \phi_0^{\prime} W(\phi_0)\,,
\ee
which furnishes the energy density
\be\label{trho}
\rho=\phi^{\prime\;2}_0-\frac{\alpha}{2}
\left(\phi_0^{\prime\;4}+3\phi^{\prime}_0 \phi^{\prime\prime}_0 W(\phi_0)\right)\,.
\ee
The first-order correction of the energy, Eq.~\eqref{E11}, can be written as
\be
E^{(1)}=\frac{1}{4}\int_{-\infty}^{\infty} dx
\,\phi^{\prime\;4}_0\,.
\ee

Now, we examine the linear stability. The transformations given by equations \eqref{Trans1} becomes
\ben
dx&=&\left(1+\alpha (n-1)W_\phi^{2n-2}\right)dz\\
\eta&=&\left(1-\frac{\alpha n}2 W_\phi^{2n-2}\right)\tilde\eta \een
The quantum-mechanical potential, as given by Eq. \eqref{ualpha}, is such that
\ben\label{ualphaw}
U_\alpha(z)&=&\left(2n(n-1)^2-1\right)W_{\phi}^{2n-2}W_{\phi\phi}^2\nonumber\\&&+(n^2-n-1)W_{\phi}^{2n-1}W_{\phi\phi\phi}\nonumber\\
&&+\frac{(1-4n+2n^2)}{2n}\Big(3W_{\phi}W_{\phi\phi}W_{\phi\phi\phi}\nonumber\\
&&+W_{\phi}^2W_{\phi\phi\phi\phi}\Big)\times \int\,dz\,W_\phi^{2n-2}\,\nonumber\\
\,
\een
where $\phi=\phi_0(z)$.

\begin{figure}[ht]
\includegraphics[scale=0.42]{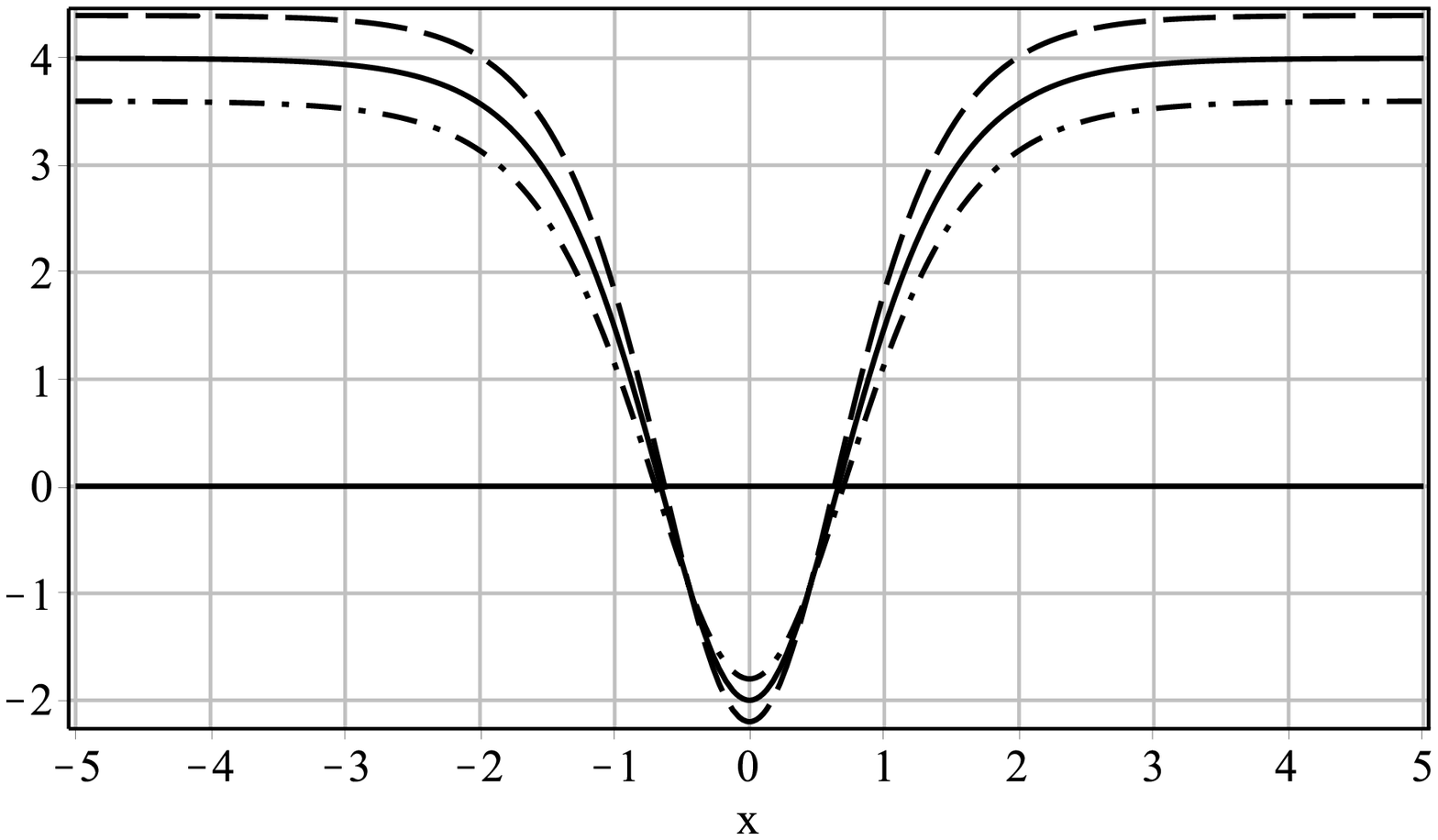}
\caption{The quantum-mechanical potential for the standard $\phi^4$ model \eqref{uphi4} (solid line), and for the generalized model, with the correction
\eqref{uphi4k1}, for $\alpha=0.1$ (dashed line) and for $\alpha=-0.1$ (dashed-dotted line).\label{figu5}}
\end{figure}

In the following, we illustrate the general case, for any integer $n$, with some examples. Let us take the $\phi^4$ model defined by the potential \eqref{phi4}. From \eqref{phia1} we obtain the first order correction of the unperturbed solution $\phi_0$ written in terms of hypergeometric function as
\ben\label{phialpha0}
\phi_\alpha&=&-\frac{2n-1}{2n}\;\tanh(x)\;\sech^2(x)\nonumber\\
&&\times{}\;_2F_1 \left(\frac12,3-2n;\frac32;\tanh(x)^2\right)\,,
\een
that represents a polynomial series of degree $4n-5$ in $\tanh(x)$. Also, from \eqref{E11}, the first-order correction to the energy is
\be 
\label{EE0} E^{(1)}=\frac{1}{2n}\frac{\Gamma(2n)
\sqrt{\pi}}{\Gamma(2n+1/2)}\,.
\ee
The correction to the quantum-mechanical potential \eqref{ualpha} can be calculated case by case. For $n=1$, we obtain the solution
\be\label{sol1}
\phi(x)=\tanh(x)-\frac12{\alpha}\;x\;\sech^2(x)\,,
\ee
the energy is given by \eqref{eny}, where $E_0=4/3$,
and the correction to the quantum-mechanical potential becomes
\be\label{uphi4k1}
U_{\alpha}(x)=-4+6\;\sech^2(x)\left(1-x\;\tanh(x)\right)\,.
\ee
It gives
$\tilde\omega_0=0$, which shows that \eqref{sol1} is stable. In Fig.~\ref{figu5}, we plot the corresponding quantum-mechanical potential.

\begin{figure}[ht]
\includegraphics[scale=0.42]{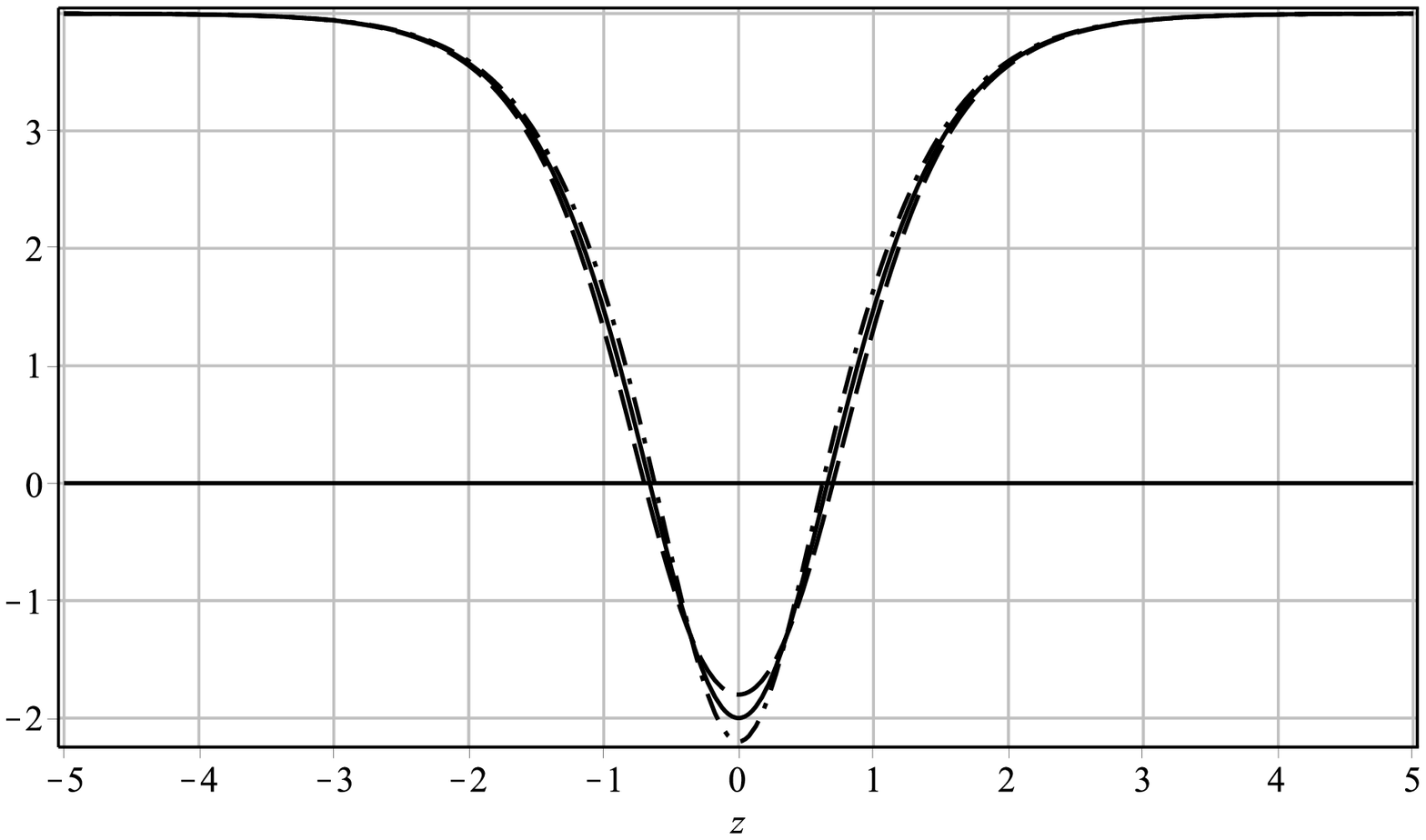}
\caption{The quantum-mechanical potential for the $\phi^4$ model \eqref{uphi4} (solid line), and for the generalized model, with the correction
\eqref{uphi4k2}, for $\alpha=0.1$ (dashed line) and for $\alpha=-0.1$ (dashed-dotted line).\label{figu6}}
\end{figure}

For $n=2$, we have
\be\label{sol2}
\phi(x)=\tanh(x)\left(1-\frac14\alpha\,\sech^2(x)\,\left(2+\sech^2(x)\right)\right)\,, 
\ee
the energy
$
E=\frac43\left(1+\frac6{35}\alpha\right)$,
and also
\be\label{uphi4k2}
U_{\alpha}(z)= 2\;\sech^2(z)+11\;\sech^4(z)-15\;\sech^6(z)
\ee
which gives
$\tilde\omega_0=0$, showing that the generalized solution \eqref{sol2} is stable. As before, in Fig.~\ref{figu6} we depict the potential for some values of $\alpha$.

Let us now apply the same procedure to the sine-Gordon model defined by \eqref{vsg}. For generic $n$, the first-order correction to the unperturbed solution \eqref{phia1} is also given by the hypergeometric function as
\ben\label{phialpha2}
\phi_\alpha&=&-\frac{2n-1}{2n}\;\tanh(x)\;\sech(x)\nonumber\\
&&\times{}\;_2F_1 \left(\frac12,2-n;\frac32;\tanh(x)^2\right)\,,
\een
and the first-order correction to the energy is
\be
\label{EE145} E^{(1)}=\frac{1}{2n}\frac{\Gamma(n)
\sqrt{\pi}}{\Gamma(n+1/2)}\,.
\ee
Again, the perturbation $U_{\alpha}$ is calculated case by case.
For $n=1$, the correction of the solution is
\be\label{sol3}
\phi_{\alpha}=-\frac12{\alpha}\;x\;\sech(x)\,,
\ee
the energy is given by \eqref{eny}, where $E_0=2$,
and the correction to the quantum-mechanical potential is
\be\label{usgk1}
U_{\alpha}(x)= -1+2\;\sech^2(x)\left(1-x\;\tanh(x)\right)\,,
\ee
giving 
$\tilde\omega_0=0$. Thus, the static solution of the generalized model is also stable.

For $n=2$, we have 
\be\label{sol4} \phi_\alpha(x)=-\frac34\tanh(x)\;\sech(x)\,,
\ee
the energy $ E=2(1+\frac16\alpha)$,
and the correction to the quantum-mechanical potential
\be\label{usgk2}
U_{\alpha}(z)=4\;\sech^2(z)-5\;\sech^4(z)\,,
\ee
which gives
$\Delta\omega_0=0$. Hence, the static solution \eqref{sol4} is also stable. 
\subsection{The case $F(\phi,X)$}

Let us now investigate the case where $F(\phi,X)= F(X)\,V_\alpha(\phi)$.
Here we can write the stress density as
\be
p_{\alpha}=(F-2F_{X}X)\; V_\alpha(\phi)\,,
\ee
and for the homogeneous and static solutions given by \eqref{Criticalpoints}, and  \eqref{pphi} with \eqref{alpha}, respectively,
we can define the effective potential
\be\label{veff}
V_{eff}= \frac12\Big(W_\phi-\alpha\frac{F(\phi,W_\phi)}{W_\phi}\Big)^2\,.
\ee
\\

We consider the kinetic perturbation given by \eqref{kpert1}, where $n$ is positive integer, and $V_\alpha=a(m,n) W_\phi^{2m}$,
where $a(m,n)=n(2m+2n-1)/(2n-1)(n+m)$, and $m$ is positive integer or semi-integer. The effective potential \eqref{veff} becomes
\be\label{veff1}
V_{eff}=\frac12\left(W_\phi+\frac{\alpha}{2n}W_\phi^{2n+2m-1}\right)^2\,.
\ee
It has the same minima and maxima of the standard potential, for $\phi_0(-\infty)\leq\phi\leq\phi_0(\infty)$. The corrections to the static solution is given by the
Eq.~\eqref{phia1}, after changing $n\rightarrow n+m$. The first-order correction of the energy, Eq.~\eqref{FirstOrder}, becomes
\be\label{E1prod}
E^{(1)}=\frac{a(m,n)}{2n}\int_{-\infty}^{\infty} dx \,\phi_0^{\prime\;2(m+n)}\,.
\ee
The correction to the quantum-mechanical potential, Eq.~\eqref{ualpha}, can be written as
\ben\label{uprod}
U_\alpha(z)&=&a(m,n)\Big\{A(m,n)W_{\phi}^{2(n+m-1)}W_{\phi\phi}^2\nonumber\\&&+B(m,n)W_{\phi}^{2(n+m)-1}W_{\phi\phi\phi}\nonumber\\
&&+\frac{(1-4n+2n^2)}{2n}\Big(3W_{\phi}W_{\phi\phi}W_{\phi\phi\phi}\nonumber\\
&&+W_{\phi}^2W_{\phi\phi\phi\phi}\Big)\times \int\,dz\,W_\phi^{2(n+m-1)}\Big\}\,\nonumber\\
\,
\een
where
\bes
\ben
A(m,n)&=&2(n-2)m^2+\left(4n^2-8n-\frac1n+4\right)m\nonumber\\
&&+2n(n-1)^2-1\,,\\
B(m,n)&=&\left(n-\frac1n-2\right)m+n^2-n-1\,,
\een\
\ees
and $\phi=\phi_0(z)$.

\begin{figure}[ht]
\includegraphics[scale=0.42]{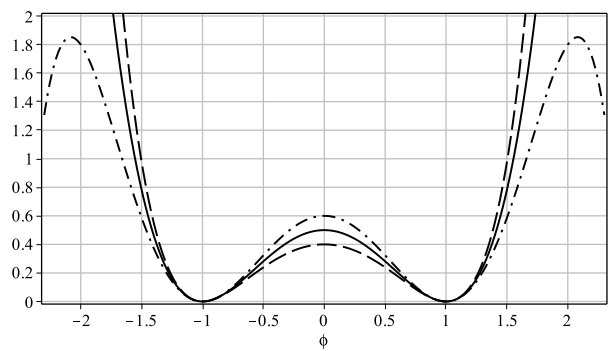}
\caption{The $\phi^4$ potential \eqref{phi4} (solid line) and the effective potential \eqref{veff} for $m=n=1$, with $\alpha=0.1$ (dashed line) and 
with $\alpha=-0.1$ (dashed-dotted line).\label{figu7}}
\end{figure}

Let us take, for example, the standard model as the $\phi^4$ potential \eqref{phi4}. The first-order correction of the static solution is given by \eqref{phialpha0}, changing $n\rightarrow n+m$, and the correction of the energy is given by
\be 
\label{E1prod1} E^{(1)}=\frac{a(m,n)}{2n}\frac{\Gamma(2n+2m)
\sqrt{\pi}}{\Gamma(2n+2m+1/2)}\,.
\ee 
For $n=m=1$, we have the solution \eqref{sol2}, with energy $E=\frac43\left(1+\frac{18}{35}\alpha\right)$, and the effective potential \eqref{veff1}, which is depicted in Fig.~\eqref{figu7}.

\begin{figure}[ht]
\includegraphics[scale=0.42]{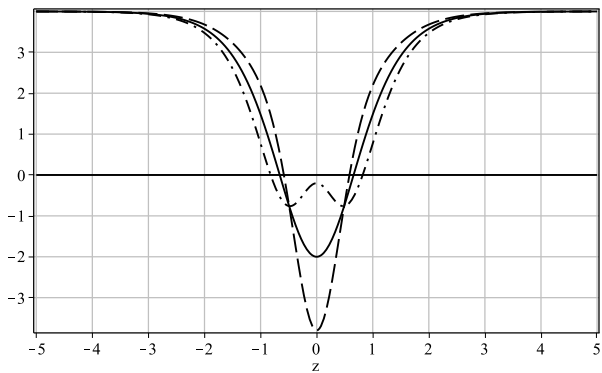}
\caption{The quantum-mechanical potential for the $\phi^4$ model \eqref{uphi4} (solid line), and the generalized quantum-mechanical potential \eqref{uprod}, for $n=m=1$ with
$\alpha=0.2$ (dashed line) and with $\alpha=-0.2$ (dashed-dotted line).\label{figu8}}
\end{figure}

The correction to the quantum-mechanical potential has the form
\be
U_{\alpha}(z)=-6\;\sech^2(z)-21\;\sech^4(z)+36\;\sech^6(z)\,,
\ee
which, from eq.\eqref{deltaw}, gives $\tilde\omega_0=343/100\;\alpha$. Thus, the defect structure is stable for positive $\alpha$; see also Fig.~\ref{figu8}.

\begin{figure}[ht]
\includegraphics[scale=0.42]{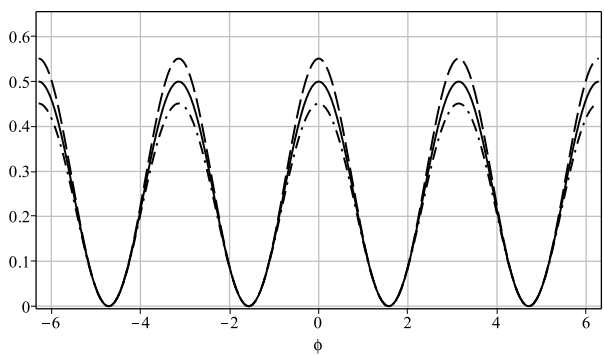}
\caption{The sine-Gordon potential \eqref{vsg} (solid line) and the effective potential \eqref{veff} for $m=n=1$, with $\alpha=0.1$ (dashed line) and with $\alpha=-0.1$, (dashed-dotted line).\label{figu9}}
\end{figure}

We now consider the standard model as the sine-Gordon model, with the potential \eqref{vsg}. The first-order correction to the static solution is given by \eqref{phialpha2}, changing $n\rightarrow n+m$, and the correction of the energy is given by
\be 
\label{E1prod2} E^{(1)}=\frac{a(m,n)}{2n}\frac{\Gamma(n+m)
\sqrt{\pi}}{\Gamma(n+m+1/2)}\,.
\ee 
For $n=m=1$, we have the solution \eqref{sol4}, with energy $E=2\left(1+\frac12\alpha\right)$, and the effective potential \eqref{veff1} is illustrated in Fig.~\ref{figu9}.

\begin{figure}[ht]
\includegraphics[scale=0.42]{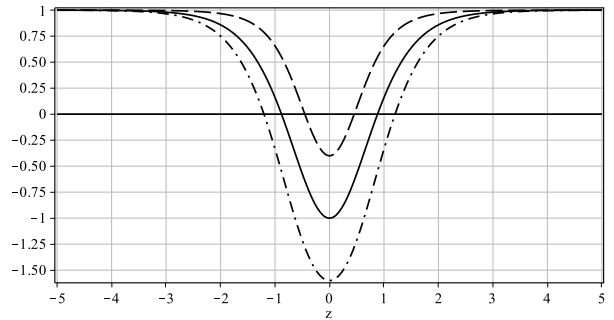}
\caption{The quantum-mechanical potential of sine-Gordon model \eqref{usg} (solid line), and the generalized quantum-mechanical potential \eqref{uprod}, for $n=m=1$, with
$\alpha=0.2$ (dashed line) and with $\alpha=-0.2$ (dashed-dotted line).\label{figu10}}
\end{figure}

The correction to the quantum-mechanical potential is
\be
U_{\alpha}(z)=-8\;\sech^2(z)+5\;\sech^4(z)\,,
\ee
which, from Eq.~\eqref{deltaw}, gives $\tilde\omega_0=109/100\;\alpha$. Again, the defect structure of the generalized model is stable for positive $\alpha$; see also Fig.~\ref{figu10}.

\section{Comments and Conclusions}

In this work, we developed a formalism to obtain and study defect structures for scalar field theory in $(1,1)$ spacetime dimensions. The models that we consider are generalized models, obtained from standard theory, with the addition of a general function
depending on the field and its derivative, in the form $F=F(X,\phi)$, for $X=\frac12\partial_\mu\phi\partial^\mu\phi$. We consider that the standard theory supports stable defect structures, and we study how to get stable structures when the addition contribution depends on a small parameter, $\alpha$, which we use to get results valid up to the first-order correction on $\alpha$.

Up to first-order in $\alpha$, we have been able to write general results for the defect structures, its energy density and energy. For the energy, we could also write results up to second-order in $\alpha$, since the second-order correction to the energy only depends on the first-order correction of the defect structures. Also, we examined linear stability,  obtaining a Schr\"odingerlike equation which makes it possible to infer stability of the defect structure from the associated quantum-mechanical potential.

To illustrate the general results, we examined three distinct extensions of the standard theory, for which we used the $\phi^4$ model, and the sine-Gordon theory. The generalizations were implemented with $F=F(\phi)$, depending only on the scalar field, $F=F(X)$, depending only on the derivative of the scalar field, and $F=F(X,\phi)$, depending on both the field and its derivative. In the first case, the defect structures are stable if they are stable in the standard theory. In the other two cases, we can stabilize the defect structures controlling the sign of the small parameter $\alpha$. In each case, several distinct examples were considered to illustrate the general results.

\acknowledgments
The authors would like to thank CAPES, CNPq and FAPESP for partial financial support.


\end{document}